\documentclass[a4paper,11pt]{article}
\pdfoutput=1
\usepackage{hyperref}
\hypersetup{colorlinks,bookmarksopen,bookmarksnumbered,
linkcolor=blus,pdfstartview=FitH,urlcolor=rossos}
\def\hhref#1{\href{http://arxiv.org/abs/#1}{#1}} 
\usepackage{multicol}
\usepackage{color}
\definecolor{rosso}{cmyk}{0,1,1,0.4}
\definecolor{rossos}{cmyk}{0,1,1,0.55}
\definecolor{rossoc}{cmyk}{0,1,1,0.2}
\definecolor{blu}{cmyk}{1,1,0,0.3}
\definecolor{blus}{cmyk}{1,0.5,0,0.5}
\definecolor{bluc}{cmyk}{1,1,0,0.1}
\definecolor{verde}{cmyk}{0.92,0,0.59,0.25}
\definecolor{verdec}{cmyk}{0.92,0,0.59,0.15}
\definecolor{verdes}{cmyk}{0.92,0,0.59,0.4}
       
\newcommand{\eq}[1]{~{\rm (\ref{eq:#1})}}

\def\circa#1{\,\raise.3ex\hbox{$#1$\kern-.75em\lower1ex\hbox{$\sim$}}\,}

\newcommand{\PR}{Phys. Rev.}

\newcommand{\beq}{\begin{equation}}
\newcommand{\eeq}{\end{equation}}

\def\circa#1{\,\raise.3ex\hbox{$#1$\kern-.75em\lower1ex\hbox{$\sim$}}\,}
\makeatletter

\def\art{\@ifnextchar[{\eart}{\oart}}
\def\eart[#1]#2#3#4#5#6{{\rm #2}, {#3 #4} {\rm (#6) #5} [{\hhref{#1}}]}
\def\hepart[#1]#2{{\rm #2, arXiv:\hhref{#1}}}
\newcommand{\oart}[5]{{\rm #1}, {#2 #3} {\rm (#5) #4}}
\newcounter{alphaequation}[equation]
\def\thealphaequation{\theequation\hbox to
0.6em{\hfil\alph{alphaequation}\hfil}}
\def\eqnsystem#1{
\def\@eqnnum{{\rm (\thealphaequation)}}
\def\@@eqncr{\let\@tempa\relax \ifcase\@eqcnt \def\@tempa{& & &} \or
  \def\@tempa{& &}\or \def\@tempa{&}\fi\@tempa
  \if@eqnsw\@eqnnum\refstepcounter{alphaequation}\fi
\global\@eqnswtrue\global\@eqcnt=0\cr}
\refstepcounter{equation} \let\@currentlabel\theequation \def\@tempb{#1}
\ifx\@tempb\empty\else\label{#1}\fi
\refstepcounter{alphaequation}
\let\@currentlabel\thealphaequation
\global\@eqnswtrue\global\@eqcnt=0 \tabskip\@centering\let\\=\@eqncr
$$\halign to \displaywidth\bgroup \@eqnsel\hskip\@centering
$\displaystyle\tabskip\z@{##}$&\global\@eqcnt\@ne
\hskip2\arraycolsep\hfil${##}$\hfil& \global\@eqcnt\tw@\hskip2\arraycolsep
$\displaystyle\tabskip\z@{##}$\hfil
\tabskip\@centering&\llap{##}\tabskip\z@\cr}
\def\endeqnsystem{\@@eqncr\egroup$$\global\@ignoretrue} \makeatother

\newcommand{\eV}{\,{\rm eV}}

\begin{document}
\twocolumn[
\begin{center}IFUP-TH/2008-10\bigskip\bigskip

{\Huge\bf\color{red}Anomalous anomalous scaling?}\\
\medskip
\bigskip\color{black}\vspace{0.6cm}
{
{\large\bf Alessandro Strumia}$^a$ and
{\large\bf Nikolaos Tetradis}$^b$
}

\bigskip
{\it  $^a$ Dipartimento di Fisica dell'Universit{\`a} di Pisa and INFN, Italia}

{\it  $^b$ Department of Physics, University of Athens, Zographou GR-15784, Athens, Greece}

\bigskip\bigskip
{\large
\centerline{\bf Abstract}
\begin{quote}{Motivated by speculations about infrared 
deviations from the standard behavior of local quantum field theories,
we explore the possibility that such effects might show up as an anomalous running of coupling constants.
The most sensitive probes are presently given by the anomalous magnetic moments of 
the electron and the muon,
that suggest that $\alpha_{\rm em}$ runs
$1.00047\pm0.00018$ times faster than predicted by the Standard Model.
The running of $\alpha_{\rm em}$ and $\alpha_{\rm s}$ up to the weak scale is
confirmed with a precision at the $\%$ level.}
\end{quote}}
\end{center}

\bigskip\bigskip
]
\normalsize

\section{Introduction}
The range of validity of Quantum Field Theory (QFT)
may be limited not only in the UltraViolet (UV), 
$E \circa{<} \Lambda_{\rm UV}$, but also in the InfraRed (IR),
$E\circa{>} \Lambda_{\rm IR}\equiv 1/L$, 
with a non trivial connection between $ \Lambda_{\rm UV}$
and $ \Lambda_{\rm IR}$.
This possibility has attracted interest due to the following reasons.

\smallskip

On the theoretical side,
requiring that the entropy associated with the QFT degrees of freedom 
$\sim (\Lambda_{\rm UV}/\Lambda_{\rm IR})^3$
saturates the Bekestein  entropy~\cite{Bek} $\sim L^2M_{\rm Pl}^2$ of a black hole 
with size
$L$ leads to $\Lambda_{\rm IR} \sim \Lambda_{\rm UV}^3/M_{\rm Pl}^2$.
Alternatively, it has been suggested that one should require that 
systems whose size $L$ exceeds their
Schwarzschild   radius $ \sim m/M_{\rm Pl}^2$
do not appear in QFT. For  $m\sim \Lambda_{\rm UV}^3 L^3$, this requirement leads 
to $\Lambda_{\rm IR}\sim \Lambda^2_{\rm UV}/M_{\rm Pl}$~\cite{Cohen}.

\smallskip

On the phenomenological side, ref.s~\cite{Cohen, Banks} discussed  possible connections of these ideas
with the cosmological constant and the supersymmetry 
breaking puzzles.
Indeed, in standard QFT the values of the vacuum energy, 
scalar masses squared and dimensionless couplings
are given by their bare Planck-scale values plus a quantum correction proportional to
$\Lambda_{\rm UV}^4$, $\Lambda_{\rm UV}^2$ and $\ln \Lambda_{\rm UV}$ respectively.
Such non-local effects could change this 
power-counting, 
solving or modifying the hierarchy problems associated with massive 
parameters~\cite{Cohen, Banks}.

We observe here that dimensionless couplings may be similarly affected, 
leading to an anomalous Renormalization Group (RG) running,
and we study how accurately present data test the standard QFT prediction.

\section{Speculations}
The above speculations about a non-local connection between the IR and UV cutoffs
do not have a very precise meaning,
and one can debate whether they would lead to any of the effects mentioned above.
Rather than arguing in any one direction, we present the uncertain issues.

Firstly, {\em when do these non-local phenomena appear in particle physics?}
The weakest possibility is only when strong gravity effects, such as 
those arising from black holes, are directly relevant. This
would practically mean never, as black hole phenomena (Hawking radiation, etc.)
are quantitatively irrelevant in all processes we can realistically observe
(possibly unless the true quantum gravity scale is much below $M_{\rm Pl}$).
The strong\-est possibility, according to which  states 
with energy $E\sim \Lambda_{\rm UV}$ that propagate
for more than $L$ do not exist and must be dropped from QFT,
contradicts experience. We see TeV $\gamma$ rays from the galactic center, 
particles with energies up to $10^{20}\eV$ from extragalactic sources, etc.

\medskip

Secondly, {\em what is the precise meaning of $\Lambda_{\rm IR}$?}
There are various possibilities, and we mention two: 
i) The IR cutoff can be defined by imposing  boundary conditions
such that QFT lives ``in a box'' with size $1/\Lambda_{\rm IR}$.
ii) $\Lambda_{\rm IR}$ is the minimal energy scale that appears in loop integrations.
In practical cases these choices can lead to very different answers.  For example,
only in the first case the IR cutoff for
the cosmological constant would be the Hubble distance $1/H$ (possibly leading to a small
$\Lambda_{\rm UV}\sim \sqrt{M_{\rm Pl} H}\sim\eV$~\cite{Cohen}). On the other hand,
$H$ nowhere appears in the one-loop correction to the vacuum energy,
equal to the value of the potential $V$ at its physical local minimum:
\beq  V \simeq V_{\rm bare} +\frac{1}{2}
 \int_{\Lambda_{\rm IR}}^{\Lambda_{\rm UV}} \frac{d^4k}{(2\pi)^4} {\rm Str} \ln ( k^2+V''_{\rm bare}).\eeq

\medskip 

If the quantum correction to the minimum of  $V$ is naturally small 
thanks to an anomalous dependence on $\Lambda_{\rm UV}(\Lambda_{\rm IR})$~\cite{Cohen}, 
one could expect a similar anomalous running of
the whole potential, and in particular of its coupling constants.
This leads to the third issue: {\em what is the precise meaning of $\Lambda_{\rm UV}$?}
In standard QFT, the one-loop corrections to any dimensionless coupling (e.g.\ the gauge boson vertex $g$) 
has the form
\beq  \label{eq:run}g(p) = 
g_{\rm bare} - \beta \frac{g^3_{\rm bare}}{8\pi} 
\left[\ln \frac{\Lambda_{\rm UV}^2}{p^2} + \hbox{finite}\right],\eeq
where $p$ is (some combination of) the external momenta that sets the IR cutoff in loop integration.
In standard QFT the physical coupling $g(p)$ `runs' with the energy $p$ of the process,
and the RG coefficient $\beta$ is a number that depends on the particle content of the theory above $p$.
(In eq.\eq{run} we assumed that all the masses are negligibly small.)

\medskip

If instead non-QFT effects produce some physical UV cutoff
 $\Lambda_{\rm UV}$ that depends on $\Lambda_{\rm IR}\sim p$,
one generically obtains an anomalous RG running of $g(p)$.
For example, in the one-loop 
approximation the running is proportional to
\beq\label{eq:delta}
 \beta \to \beta 
\left( 1 - \frac{\partial \ln \Lambda_{\rm UV}}{\partial \ln \Lambda_{\rm IR}}\right) 
\equiv \beta (1-\delta).\eeq
Even so, one could argue that no anomalous running $ \delta$ needs to appear, because
$g_{\rm bare}$ might depend on $\Lambda_{\rm UV}$ in a way that counter-acts the 
explicit dependence on $\Lambda_{\rm UV}$ in eq.\eq{run}~\cite{Cohen}.
A physical realization of this mathematical possibility is that
the Standard Model (SM) at high energies below the Planck mass gets replaced by some other model
where couplings do not run (e.g.\ an UV-finite theory, or some fixed point of the RG flow).

\medskip

By using the $\beta$ function of eq.\eq{delta} within  the dimensional 
regularization formalism, we get
for the 
one-loop RG running of a gauge coupling $\alpha$
\beq\label{eq:run1}
\frac{1}{\alpha(\mu')} - \frac{1}{\alpha(\mu)} \simeq  \beta \ln \frac{\mu^{1 - \delta}}{\mu^{\prime 1-  \delta}} + \cdots ~ .\eeq
This is equivalent to the standard expression, 
with the Minimal Subtraction mass scale $\mu$ replaced by
$ \mu^{1-\delta}$.
In theories with several particle masses and sizable RG corrections, 
the factor $\delta$ generically can become
some unknown function of the energy. In order 
to compute it, we would need to know the physics around $\Lambda_{\rm UV}$.

In the next sections, we explore
the most sensitive experimental probes of 
anomalous RG running,
assuming for definiteness  
that all Standard Model formul\ae{} get modified as in eq.\eq{run},
with a constant $\delta$ to be extracted from data.



%


\section{Running of $\alpha_{\rm em}$ from $m_e$ to $m_\mu$}
The measurements of the anomalous magnetic moments of the electron~\cite{ge} 
\beq g_e/2 = 1.00115965218085(76)\eeq
and of the muon~\cite{gmu}
\beq
g_\mu/2 = 1.00116592080(63),\eeq
together with the assumption of the validity of the Standard Model,
allow us to infer the electromagnetic coupling $\alpha_{\rm em}(\mu)$ 
at the scales
$\mu=m_e$ and $m_\mu$, 
in view of the theoretical prediction
\beq g_i=2+\alpha_{\rm em}(m_i)/\pi +\cdots,\eeq
where $\cdots$ denotes higher-order effects.
We recall that $g_e$ gives the most precise determination of $\alpha_{\rm em}$,
that is consistent with lower-energy probes from atomic physics~\cite{ge}.
Assuming the anomalous running
\beq\frac{1}{\alpha_{\rm em}(m_e)} - \frac{1}{\alpha_{\rm em}(m_\mu)} =  \frac{1-\delta}{3\pi} \ln \frac{m_\mu}{m_e} + \cdots\eeq
one gets the presently most precise determination of $\delta$:
\beq \delta= -(0.047\pm 0.018)~\%   .\eeq
The central value of $\delta$ is about $3\sigma$ below zero,
because,  for $\delta=0$, $g_\mu$ is about $3\sigma$ above the SM prediction,
$(g_\mu  - g_\mu^{\rm SM})/2 = (23\pm 9)\cdot 10^{-10}$,
with the precise number depending on how one deals with the theoretical uncertainties
on higher-order QCD corrections to $g_\mu$:
relying on $e^-e^+$ data and/or on $\tau$-decay data~\cite{gmu}.

The usual new-physics interpretation of the $g_\mu-2$ anomaly is that 
new particles with heavy mass $M$, like supersymmetric particles, 
affect $g_\mu$ giving an extra contribution $\Delta g_\mu \sim \alpha_2 m_\mu^2/ M^2$.
They also
affect precision data at higher energies, but have a negligible influence on $g_e$
in view of $m_\mu \gg m_e$.

We point out that the relative incompatibility between $g_\mu$ and $g_e$
could instead be due to a `too fast' RG running of $\alpha_{\rm em}$.
We show that $g_e$ and $g_\mu$ presently 
give the most sensitive  probes to $\delta$:
this kind of new physics is best seen with higher precision than with higher energy.


\section{Running of $\alpha_{\rm em}$ from $m_\mu$ to $M_Z$}
Precision tests at the $Z$ pole offer another precision determination of the electromagnetic coupling.
By performing a global fit within the SM with Higgs mass $m_h$~\cite{EWPT} we find
\beq \frac{1}{\alpha_{\rm em}(M_Z)} = 128.92+0.23\ln \frac{m_h}{M_Z}\pm 0.06.\eeq
This value can be compared with the RG extrapolation from $m_e,m_\mu$ up to $M_Z$~\cite{had}
\beq \frac{1}{\alpha_{\rm em}(M_Z)} = 128.937+8.1\delta \pm 0.028, \eeq
where the uncertainty comes from QCD thresholds.
So 
\beq \delta=\left( - 0.2+2.9\ln\frac{m_h}{M_Z}\pm  0.9 \right)~\% .\eeq
The precise measurement of the muon lifetime does not give another probe of $\delta$, 
as the anomalous dimension of the associated Fermi operator 
$$[\bar\mu \gamma_\mu P_L \nu_\mu][\bar\nu_e  \gamma_\mu P_L  e]$$
is zero: indeed the electromagnetic current is not renormalized, and this operator
can be related to it, times a neutrino current not affected by electromagnetic interactions.

\section{Running of $\alpha_{\rm s}$ from $m_\tau$ to $m_Z$}
Another sensitive probe to $\delta$ comes from the running of the strong coupling $\alpha_{\rm s}$:
in view of its large value, $\alpha_{\rm s}$ runs fast.
The strong coupling constant has been measured at various scales,
and the two most precise determinations are at $m_\tau$ and $M_Z$.
By performing a global fit of electroweak precision data
within the SM with Higgs mass $m_h$~\cite{EWPT} we find
\beq {\alpha_{\rm s}(M_Z)} = 0.121+0.0008\ln \frac{m_h}{M_Z}\pm 0.0023.\eeq
On the other hand, the measurement of the strong coupling from $\tau$ decays,
$\alpha_{\rm s}(m_\tau)=0.334\pm 0.009$, extrapolated up to $M_Z$ gives~\cite{alpha3}
\beq {\alpha_{\rm s}(M_Z)} = 0.1212+0.08\, \delta \pm 0.0011.\eeq
So
\beq \delta = \left( -0.4+1.1 \ln\frac{m_h}{M_Z} \pm 3.3 \right)\%.\eeq
Finally, flavor-physics observations allow us to test the QCD running of various operators from the weak scale
down to the bottom or charm mass. However, the uncertainty on $\delta$ is at the level of
several tens of percent.

\section{Conclusions}
Motivated by possible deviations from the standard QFT predictions for the RG running of
couplings, 
we rescaled $\beta$ functions by $1-\delta$ and studied how data probe the new-physics 
parameter $\delta$
that parameterizes an anomalous running.
Unlike in ordinary new physics, the most sensitive probe to $\delta$ is given by
precision experiments at low energies $E\circa{>} m_e$:
the measurements of the magnetic moments of the electron and the muon
determine $\delta$ with a $0.018\%$ uncertainty,
excluding order-one effects.
However, the anomaly in the anomalous magnetic moment of  the muon
indicates a best fit value for $\delta$ which is $3\sigma$ below zero.
Running of $\alpha_{\rm em}$ and $\alpha_{\rm s}$ 
up to $M_Z$ is confirmed with a $1\%$ and $3\%$ precision respectively.

\paragraph{Acknowledgements}
We thank Gino Isidori for  comments.

\end{document}